\documentclass[letter]{ieice}
\usepackage{latexsym}

\field{A}
\vol{91}
\no{10}
\SpecialSection{Information Theory and its Applications}
\title[Mismatched measurements in the BB84 protocol]{Key Rate Available from Mismatched Mesurements in the
BB84 Protocol and the Uncertainty Principle}
\titlenote{To appear in IEICE Trans.\ Fundamentals (http://ietfec.\\ oxfordjournals.org/), vol.\ E91-A, no.\ 10, Oct.\ 2008. The page number has not been determined yet.}
\authorlist{
 \authorentry[ryutaroh@rmatsumoto.org]{Ryutaroh MATSUMOTO}{m}{titech}
 \authorentry{Shun WATANABE}{s}{titech}
}
\affiliate[titech]{The authors are with the Department of Communications and
Integrated Systems, Tokyo Institute of Technology, Tokyo, 152-8550 Japan.}

\received{2008}{1}{15}
\revised{2008}{3}{31}

\newcommand{\ket}[1]{|#1\rangle}
\newcommand{\bra}[1]{\langle #1 |}

\begin{document}
\maketitle
\begin{summary}
We consider the mismatched measurements in the
BB84 quantum key distribution protocol,
in which measuring bases are different from transmitting bases.
We give a lower bound on the amount of a secret key that can be
extracted from the mismatched measurements.
Our lower bound shows that we can extract a secret key
from the mismatched measurements with certain quantum channels,
such as the channel over which the Hadamard matrix is applied to
each qubit with high probability.
Moreover,
the entropic uncertainty principle
implies that one cannot extract the secret key from
both matched measurements and mismatched ones simultaneously,
when we use the standard information reconciliation and
privacy amplification procedure.
\end{summary}

\section{Introduction}
The BB84 protocol \cite{bennett84}
is one of the most-known protocols for
quantum key distribution.
In this protocol,
the sender, Alice, sends qubits in one of four
quantum state vectors $\ket{0}$, $\ket{1}$,
$\ket{+} = (\ket{0}+\ket{1})/\sqrt{2}$,
$\ket{-} = (\ket{0}-\ket{1})/\sqrt{2}$,
where $\{ \ket{0}$, $\ket{1}\}$ forms an orthonormal basis.
Then the receiver, Bob, measures the received qubits
with either $\{\ket{0}$, $\ket{1}\}$ or $\{\ket{+}$, $\ket{-}\}$
basis.
After that,
Alice publicly announces to which $\{\ket{0}$, $\ket{1}\}$ or $\{\ket{+}$, $\ket{-}\}$ basis each qubit belong.
Bob discard the measurement outcomes whose bases do not contain
the transmitted qubit.
We call such measurement \emph{mismatched measurement}
in this paper.
After that, Alice and Bob perform the information reconciliation
and the privacy amplification to obtain the same secret key
as described in \cite{shor00}.

As far as the authors know,
there is no literature that clarifies the amount of
key that can be extracted from mismatched measurements in the BB84 protocol,
though Pawlowski \cite{pawlowski07} studied the same problem
with a protocol  completely different from the BB84.
The efficiency of a quantum key distribution (QKD) protocol
is measured by the ratio of extracted bits of secret key per
remaining bits not disclosed nor discarded,
which is called key rate.
We shall show a lower bound
on the key rate 
that can be extracted from mismatched measurements in the BB84 protocol.
Moreover, we shall show that,
when we use the standard information reconciliation and privacy amplification
in \cite{shor00} and the well-known lower bound
on the key rate \cite{gottesman03},
the entropic uncertainty principle \cite{maassen88}
implies that one cannot extract the secret key from
both matched measurements and mismatched ones simultaneously.

The reader may think that the probability of
getting the measurement outcome $\ket{+}$ is always $0.5$
when the transmitted qubit is $\ket{0}$,
and considering mismatched measurement outcomes is useless
and nonsense.
We cannot extract a secret key
from mismatched measurement outcomes when such a probability is $0.5$.
However, consider the quantum channel over which the Hadamard matrix
is applied to each qubit with high probability.
With such a quantum channel the above probability is close to
$0$ or $1$. The lower bound obtained in this paper
shows that we can extract a secret key from mismatched measurement
outcomes with such a quantum channel.

This paper is organized as follows:
Section~\ref{sec2} presents a variant of the BB84 protocol.
Section~\ref{sec3} verifies its unconditional security
and shows a lower bound on its key rate.
Section~\ref{sec4}
shows that the lower bounds on the key rates available
from matched measurements and mismatched ones cannot
be simultaneously positive,
by using the entropic uncertainty principle \cite{maassen88}.
Section~\ref{sec5}
gives concluding remarks and lists several open research
questions.

\section{Protocol}\label{sec2}
In this section,
we shall show a variant of the BB84 protocol
that tries to extract secret key from mismatched measurement outcomes.
We define the matrices $X$ and $Z$ representing
the bit error and the phase error, respectively, as
\begin{eqnarray*}
&&X\ket{0} = \ket{1},\quad X\ket{1}=\ket{0},\\
&&Z\ket{+} = \ket{-},\quad Z\ket{-}=\ket{+}.
\end{eqnarray*}

\begin{enumerate}
\item\label{l1}
Alice makes a random qubit sequence according to
the i.i.d.\ uniform distribution on $\{\ket{0}$, $\ket{1}$,
$\ket{+}$, $\ket{-}\}$ and sends it to Bob.
\item\label{l2}
Bob chooses the $\{\ket{0}$, $\ket{1}\}$ basis or $\{\ket{+}$, $\ket{-}\}$
basis uniformly randomly for each received qubit and measure it
by the chosen basis.
\item\label{l4}
Alice publicly announces which basis $\{\ket{0}$, $\ket{1}\}$ or
$\{\ket{+}$, $\ket{-}\}$ each transmitted qubit belongs to.
Bob also publicly announces which basis was used for measurement
of each qubit.
In the following steps they will only consider qubits with which
transmission basis and measuring bases \emph{do not} coincide between
Alice and Bob.
\item\label{l5}
Suppose that there are $2n$ qubits transmitted in the $\{\ket{0}$, $\ket{1}\}$
basis and measured with the $\{\ket{+}$, $\ket{-}\}$ basis by Bob.
Index those qubits by $1$, \ldots, $2n$.
Define the bit $a_i = 0$ if Alice's $i$-th qubit was $\ket{0}$,
and $a_i=1$ otherwise.
Define the bit $b_i = 0$ if Bob's measurement outcome
for $i$-th qubit was $\ket{+}$,
and $b_i=1$ otherwise.
\item\label{l6}
Suppose also that there are $2n'$ qubits transmitted in the $\{\ket{+}$, $\ket{-}\}$
basis and measured with the $\{\ket{0}$, $\ket{1}\}$ basis by Bob.
Index those qubits by $1$, \ldots, $2n'$.
Define the bit $\alpha_i = 0$ if Alice's $i$-th qubit was $\ket{+}$,
and $\alpha_i=1$ otherwise.
Define the bit $\beta_i = 0$ if Bob's measurement outcome
for $i$-th qubit was $\ket{0}$,
and $\beta_i=1$ otherwise.
\end{enumerate}
For the simplicity of the presentation,
we shall describe the procedure extracting the secret key
from $a_i$ and $b_i$.
The key rate for $\alpha_i$, $\beta_i$ will turn out to be
the same as that for $a_i$, $b_i$ at the end of Section~\ref{sec32}.
In the following steps,
the half of measurement outcomes will be disclosed
and used for
estimation of error rates $q_X$ and $q_Z$.
The amount of disclosure can be arbitrarily chosen
provided that the estimation of error rates can be done
sufficiently accurately.
\begin{enumerate}
\setcounter{enumi}{5}
\item\label{l7} Alice chooses a subset $S \subset \{1$, \ldots, $2n\}$
with size $|S| = n$ uniformly randomly from subsets of 
$\{1$, \ldots, $2n\}$, and publicly announces the choice of $S$.
Alice and Bob publicly announce $a_i$ and $b_i$
for $i\in S$ and compute the error rate
\begin{equation}
q_{X} = \frac{|\{i\in S \mid a_i \neq b_i \}|}{|S|}. \label{eq:q1}
\end{equation}
\item\label{l8} Alice chooses a subset $S' \subset \{1$, \ldots, $2n'\}$
with size $|S'| = n'$ uniformly randomly from subsets of 
$\{1$, \ldots, $2n'\}$, and publicly announces the choice of $S'$.
Alice and Bob publicly announce $\alpha_i$ and $\beta_i$
for $i\in S'$ and compute the error rate
\begin{equation}
q_{Z} = 
\frac{|\{i \in S' \mid \alpha_i \neq \beta_i\}|}{|S'|}. \label{eq:q2}
\end{equation}
\item\label{l9}
Alice and Bob decide a linear code $C_1$ of length $n$
such that its decoding error probability is sufficiently small
over all the binary symmetric channel whose crossover probability
is close to $q_X$.
Let $H_1$ be a parity check matrix for $C_1$,
$\vec{a}$ be Alice's remaining (not announced) bits among $a_i$'s,
and $\vec{b}$ be Bob's remaining bits among $b_i$'s.
\item\label{l10}
Alice publicly announces the syndrome $H_1\vec{a}$.
\item\label{l11}
If $q_X > 0.5$ then Bob negates every bit in $\vec{b}$
before executing the following steps.
\item\label{l12}
Bob compute the error vector $\vec{f}$ such that
$H_1\vec{f}=H_1\vec{b}-H_1\vec{a}$ by the decoding
algorithm for $C_1$. With high probability $\vec{b}-\vec{f} = \vec{a}$.
\item\label{l13}
Alice chooses a subspace $C_2 \subset C_1$ with $\dim C_2 = nh(q_Z)$
uniformly randomly, where $h$ denotes the binary entropy function,
and publicly announces her choice of $C_2$.
The final shared secret key is the coset $\vec{a}+C_2$.
\end{enumerate}

When measuring bases are the same as the transmitting bases,
we can use the standard BB84 protocol. Thus, we discard no
measurement outcome when we combine the above protocol
with the standard BB84.

\section{Security proof and a lower bound on the key rate}\label{sec3}
We shall verify the unconditional security of our proposed
protocol by directly relating it to the quantum error correction by
the quantum CSS (Calderbank-Shor-Steane) codes \cite{calderbank96,steane96}.
To make this paper self-contained, we shall briefly review the
CSS code.
After that
we shall relate our variant of the BB84 protocol to the CSS code
in a similar way to Shor and
Preskill \cite{shor00}.

\subsection{Review of the CSS code}
For a binary vector
$\vec{v} = (v_1$, \ldots, $v_n) \in \mathbf{F}_2^n$,
where $\mathbf{F}_2$ is the Galois field with two elements,
we define the quantum state vector $\ket{\vec{v}}$ by
\[
\ket{\vec{v}} = \ket{v_1} \otimes  \cdots \otimes
\ket{v_n}.
\]
For two binary linear codes $C_2 \subset C_1 \subset \mathbf{F}_2^n$,
the CSS code is the complex linear space spanned by the vectors
\[
\frac{1}{\sqrt{|C_2}|}
\sum_{\vec{w} \in C_2} \ket{\vec{v}+\vec{w}}, 
\]
for all $\vec{v}\in C_1$.
We also need parameterized CSS codes introduced in
\cite{shor00}.
The parameterized CSS code for $\vec{x},\vec{z}\in\mathbf{F}_2^n$
is defined as the linear space spanned by
\begin{equation}
\frac{1}{\sqrt{|C_2}|}
\sum_{\vec{w} \in C_2} (-1)^{(\vec{z},\vec{w})}\ket{\vec{x}+\vec{v}+\vec{w}}, \label{css2}
\end{equation}
for all $\vec{v}\in C_1$,
where $(\cdot,\cdot)$ denotes the inner product.

\subsection{Security proof and analysis of the key rate}\label{sec32}
We shall first show that our protocol is equivalent to
sending a parameterized CSS codeword with the parameter $\vec{z}$ randomly chosen.
If we fix $\vec{v}$ and $\vec{x}$ and choose $\vec{z}$ uniformly
randomly in Eq.~(\ref{css2}), then the resulting density operator
is
\begin{eqnarray}
&& \frac{1}{2^n|C_2|}\sum_{\vec{z}\in\mathbf{F}_2^n}
\left(
\sum_{\vec{w_1} \in C_2} (-1)^{(\vec{z},\vec{w_1})}\ket{\vec{x}+\vec{v}+\vec{w_1}}\right)\nonumber\\
&&
\left(
\sum_{\vec{w_2} \in C_2} (-1)^{(\vec{z},\vec{w_2})}\bra{\vec{x}+\vec{v}+\vec{w_2}}\right)\nonumber\\
&=& \frac{1}{|C_2|}
\sum_{\vec{w}\in C_2} \ket{\vec{x}+\vec{v}+\vec{w}}\bra{\vec{x}+\vec{v}+\vec{w}}, \label{eq:mix0}
\end{eqnarray}
by the exactly same argument as \cite{shor00}.

Denote the right hand side of Eq.~(\ref{eq:mix0})
by $\rho(\vec{x},\vec{v})$.
By a straightforward computation we can see
\begin{equation}
\frac{1}{2^n|C_1|}
\sum_{\vec{x}\in\mathbf{F}_2^n}\sum_{\vec{v}\in C_1}
\rho(\vec{x},\vec{v})
=
\frac{1}{2^n}\sum_{\vec{a}\in\mathbf{F}_2^n}\ket{\vec{a}}\bra{\vec{a}}.
\label{eq:mix}
\end{equation}
The right hand side of Eq.~(\ref{eq:mix})
means sending $\ket{0}$ or $\ket{1}$ $n$ times with equal probability,
which is exactly what Alice is doing in our protocol.
Announcing the syndrome $H_1\vec{a}$ in Step~\ref{l10}
is equivalent to announcing which $\vec{x}$ is chosen.

Consider the Hadamard matrix $H$ defined by
$H \ket{0} = \ket{+}$ and $H\ket{1} = \ket{-}$.
Consider the memoryless quantum channel $\Gamma$
over which the error $H X$ occurs with probability
$r_X$, $H Z$ occurs with probability $r_Z$, 
$H XZ$ occurs with probability $r_{XZ}$,
and $H$ occurs with probability $1-r_X-r_Z-r_{XZ}$,
with $q_X = r_X + r_{XZ}$ and $q_Z = r_{Z} + r_{XZ}$.
The qubits received by Bob
can be regarded as the output of $\Gamma$
when Alice sends $\ket{0}$ and $\ket{1}$ with equal probability,
which is equivalent to sending a quantum codeword in the CSS code
as described above.

If we apply $H^{-1}$ to each qubit,
then the quantum channel can be regarded as
causing the $X$ error with probability $r_X$,
the $Z$ error with probability $r_Z$, and
the $XZ$ error with probability $r_{XZ}$.
Therefore, if we use the standard decoding procedure
of the CSS code after applying $H^{-1}$ to each qubit in
the received codeword, then the transmitted CSS codeword
is recovered with high fidelity provided that $q_X < 0.5$ and
$q_Z < 0.5$.
Observe also that this imaginary quantum decoding process
is equivalent to what is actually performed in our
variant of the BB84 protocol described in Section~\ref{sec2}
by the almost same argument as \cite{shor00}.

When $q_X > 0.5$ and $q_Z < 0.5$,
applying $X$ to each qubit in the received codeword
after $H^{-1}$
makes $q_X$ to $1-q_X$ and leaves $q_Z$ unchanged.
When $q_X < 0.5$ and $q_Z > 0.5$, apply $Z$ to each qubit,
and when $q_X > 0.5$ and $q_Z > 0.5$, apply $XZ$.
By the above operations we can regard both $q_X$ and $q_Z$ being
less than $0.5$ provided that $q_X \neq 0.5$ and $q_Z \neq 0.5$.
Observe that application of $Z$ is purely imaginary and does not
correspond to the actual operations in the protocol
in Section~\ref{sec2}.
On the other hand,
application of $X$ corresponds to flipping each bit
in Step~\ref{l11} in our protocol.
Application of $XZ$ is equivalent to that of $X$ in
the actual protocol.

We have shown that the procedure in Section~\ref{sec2}
can be regarded as the quantum error correction of the CSS
code over a peculiar quantum channel $\Gamma$.
It was shown in Corollary 2 of \cite{csiszar82}
that there exists a linear code $C_1$ of information rate
$1-h(q_X)$ satisfying the condition in Step~\ref{l9}.
If we use such $C_1$ then we can correct $H X$ errors on $\Gamma$
with high probability.

It is stated in \cite{shor00} and proved in \cite{watanabe06} that
random choice of [$n-nh(q_Z)$]-dimensional subspace $C_2$ in $C_1$
almost always gives the low phase error decoding probability
in the standard CSS decoding procedure.
This implies that 
randomly chosen [$n-nh(q_Z)$]-dimensional subspace $C_2$ in $C_1$
can almost always correct $H Z$ errors on $\Gamma$.
Therefore,
if the choice of $C_1$ is appropriate,
then the fidelity of quantum error correction in the
imaginary transmission of the CSS codeword (\ref{css2}) over the channel
$\Gamma$
is close to $1$,
which implies that the eavesdropper Eve can obtain little
information by the same argument as \cite{hamada-qkd},
which shows the security of the BB84 protocol
directly relating it to the quantum error correction
without use of entanglement distillation argument.

Therefore, we can extract $1-h(q_X)-h(q_Z)$ bit of secret key
from one bit of the raw bits $\vec{a}$.
With a similar argument,
we can also see that the key rate available from $\alpha_i$'s
is $1-h(q_X)-h(q_Z)$.
Because with the key rate from $\alpha_i$'s
the roles of $q_X$ and $q_Z$ are interchanged,
which does not change the key rate $1-h(q_X)-h(q_Z)$.

\section{Implication by the uncertainty principle}\label{sec4}
At the end of Section~\ref{sec2},
we stated that we try to extract a secret key from both
matched measurement outcomes and mismatched ones.
In this section,
we shall consider the relation between the amount of secret key
extracted from matched measurement outcomes and that from
mismatched ones.
We shall show that
we cannot extract secret key from both matched and
mismatched measurement outcomes
by the entropic version \cite{maassen88} of the uncertainty principle,
when we use the information reconciliation and privacy amplification
in \cite{shor00} and the lower bound on its key rate in \cite{gottesman03}.
Note that Koashi already used the
uncertainty principle for security analysis of QKD protocols
\cite{koashi06}.

Maassen and Uffink \cite{maassen88} (see also
Box 11.1 of \cite{chuangnielsen})
proved the following version of uncertainty principle
in terms of the Shannon entropy.
Let $\rho$ be a density operator of a qubit.
Let $P_{01}$ (resp.\  $P_{+-}$)
be the probability distribution of
the measurement outcome by measuring $\rho$ by $\{\ket{0}$, $\ket{1}\}$ 
(resp.\ $\{\ket{+}$, $\ket{-}\}$).
Let $H(\cdot)$ denotes the Shannon entropy.
Then we have $H(P_{01}) + H(P_{+-}) \geq 1$ as described at
Eq.~(11.3) in \cite{chuangnielsen},
where the entropy is counted in the unit of bits.

Let $\Gamma$ be the memoryless quantum channel
that represents Eve's manipulation and
the channel noise between Alice and Bob.
$\Gamma$ is a map between density operators.
Define
\begin{eqnarray*}
p_{X-} &=& \bra{+} \Gamma(\ket{-}\bra{-}) \ket{+},\\
p_{X+} &=& \bra{-} \Gamma(\ket{+}\bra{+}) \ket{-},\\
q_{X1} &=& \bra{+} \Gamma(\ket{1}\bra{1}) \ket{+},\\
q_{X0} &=& \bra{-} \Gamma(\ket{0}\bra{0}) \ket{-},\\
p_{Z1} &=& \bra{0} \Gamma(\ket{1}\bra{1}) \ket{0},\\
p_{Z0} &=& \bra{1} \Gamma(\ket{0}\bra{0}) \ket{1},\\
q_{Z-} &=& \bra{0} \Gamma(\ket{-}\bra{-}) \ket{0},\\
q_{Z+} &=& \bra{1} \Gamma(\ket{+}\bra{+}) \ket{1}.
\end{eqnarray*}
Observe that $q_X = (q_{X0}+q_{X1})/2$
and $q_Z = (q_{Z+}+q_{Z-})/2$,
where $q_X$ and $q_Z$ are as defined in Eqs.~(\ref{eq:q1}) and
(\ref{eq:q2}).

Define $p_X = (p_{X+}+p_{X-})/2$ and 
$p_Z = (p_{Z0}+p_{Z1})/2$.
Observe that $p_Z$ (resp.\ $p_X$)
is the error rate of the matched measurement
in the BB84 protocol when transmitting basis is $\{\ket{0}$,
$\ket{1}\}$ (resp.\ $\{\ket{+}$, $\ket{-}\}$).
The lower bound on key rate of the plain one-way postprocessing described
in \cite{shor00} is given by $1-h(p_X)-h(p_Z)$ \cite{gottesman03}.

{From} the entropic uncertainty principle reviewed at the beginning
of this section, we have
\begin{eqnarray}
h(p_{X+}) + h(q_{Z+}) &\geq&1,\label{f1}\\
h(p_{X-}) + h(q_{Z-}) &\geq&1,\label{f2}\\
h(p_{Z0}) + h(q_{X0}) &\geq&1,\label{f3}\\
h(p_{Z1}) + h(q_{X1}) &\geq&1.\label{f4}
\end{eqnarray}
By the concavity of the entropy function
\begin{eqnarray}
\hspace*{-5mm}h(p_X) = h\left(\frac{p_{X+}+p_{X-}}{2}\right) &\geq&\frac{h(p_{X+})+h(p_{X-})}{2},\label{f5}\\
\hspace*{-5mm}h(q_Z) = h\left(\frac{q_{Z+}+q_{Z-}}{2}\right) &\geq&\frac{h(q_{Z+})+h(q_{Z-})}{2}\label{f8},\\
\hspace*{-5mm}h(p_Z) = h\left(\frac{p_{Z0}+p_{Z1}}{2}\right) &\geq&\frac{h(p_{Z0})+h(p_{Z1})}{2},\label{f6}\\
\hspace*{-5mm}h(q_X) = h\left(\frac{q_{X0}+q_{X1}}{2}\right) &\geq&\frac{h(q_{X0})+h(q_{X1})}{2}.\label{f7}
\end{eqnarray}
Applying Eqs.~(\ref{f5}) and (\ref{f8}) to the sum of Eqs.~(\ref{f1}) and
(\ref{f2}) divided by two, we obtain
\begin{equation}
h(p_X) + h(q_Z) \geq 1. \label{f9}
\end{equation}
{From} Eqs.~(\ref{f3}, \ref{f4}, \ref{f6}, \ref{f7}) in a similar manner
we obtain
\begin{equation}
h(p_Z) + h(q_X) \geq 1. \label{f10}
\end{equation}
Equations (\ref{f9}) and (\ref{f10}) imply
\begin{equation}
\hspace*{-3mm}[1 - h(p_X) - h(p_Z)] + [1 - h(q_X) - h(q_Z)] \leq 0. \label{f11}
\end{equation}
The first term in Eq.~(\ref{f11}) is the lower bound on key rate of the
matched measurement in the BB84 protocol,
while the second term is that of the mismatched measurement.
Equation (\ref{f11}) means that both lower bounds
on key rates cannot be simultaneously
positive.

\section{Concluding remarks}\label{sec5}
We proposed a variant of the BB84 protocol
that extracts secret key from measurement outcomes
with which measuring bases are different from transmitting bases,
obtained a lower bound on the key rate which has a similar form
to that of the standard BB84 protocol, and verified its
unconditional security.
After that, we showed that the lower bounds on the key rates
available from matched and mismatched measurements
cannot be simultaneously positive.

Our result generates a number of interesting research
questions. Firstly, the well-known upper bound (Table I in
\cite{gottesman03}) on the key rates
of the BB84 protocol \cite{shor00} is expressed
in terms of error rates for matched measurements with
which measuring bases are the same as the transmitting bases.
If both error rates for the matched bases are $0.5$,
for example the Hadamard matrix is applied to every qubit,
then those upper bounds state that we cannot extract secret key.
However, our proposed variant of the BB84 protocol can extract
secret key in such case. It is desirable to have an upper bound
on the amount of available secret key
taking into account mismatched measurement in the BB84 protocol.

Secondly, we could not disprove the possibility that
we can extract secret key from both matched and mismatched
measurements by more sophisticated postprocessing of the
BB84 protocol such as Refs.~\cite{gottesman03,renner05,ma06,watanabe07}.
It is desirable to prove or disprove such possibility.

Thirdly,
in the standard BB84 protocol,
it is generally believed that
we can postprocess bits transmitted by $\{\ket{0}$,
$\ket{1}\}$ basis and $\{\ket{+}$, $\ket{-}\}$ basis
separately without decreasing the key rate.
The proposed variant of the BB84 protocol
separately postprocess bits obtained by matched measurements
and mismatched ones, because the matched measurement outcomes
are processed with the standard BB84 protocol separately.
It is not clear whether or not such separate processing
of matched and mismatched measurement outcomes
decreases the total amount of secret key.
We leave these questions as future research agenda.

\section*{Acknowledgment}
This research was partly supported
by the Japan Society for the Promotion of Science
under Grants-in-Aid No.\ 18760266 and
No.\ 00197137.


\end{document}